\documentclass[12pt,preprint]{aastex}

\def\deg{^\textrm{\scriptsize o}}

\shorttitle{IRAS 16552$-$3050: a new ``water fountain''}
\shortauthors{Su\'arez et al.}

\begin{document}

\title{VLA observations of the ``water fountain'' IRAS\,16552$-$3050}

\author{Olga  Su\'arez\altaffilmark{1,2}, Jos\'e  F. G\'omez\altaffilmark{2}, 
Luis F. Miranda\altaffilmark{2}}
\altaffiltext{1}{UMR 6525 H.Fizeau, Univ. Nice Sophia Antipolis, CNRS, OCA. Parc
  Valrose, F-06108 Nice cedex 02, France. olga.suarez@unice.fr}

\altaffiltext{2}{Instituto de Astrof\'{\i}sica de Andaluc\'{\i}a, CSIC, Apartado
  3004, E-18080 Granada, Spain}

\begin{abstract}

We present Very Large Array (VLA) observations of the water maser
emission towards IRAS 16552-3050. The maser emission shows a
velocity spread of $\simeq 170$ km~s$^{-1}$, and a bipolar
distribution with a separation between the red and blueshifted
groups of $\simeq 0.08''$. These observations and the likely
post-AGB nature of the source indicate that IRAS 16552-3050 can be
considered as a member of the ``water fountain" class of sources
(evolved stars showing H$_2$O maser emission with a velocity spread
$\ga 100$ km~s$^{-1}$, probably tracing collimated jets).  The water
maser emission in IRAS 16552-3050 does not seem to be associated
with with any known optical counterpart. Moreover, this source does
not have a near-IR 2MASS counterpart, as it happens in about half of
the water fountains known. This suggests that these sources tend to
be heavily obscured objects, probably with massive precursors ($\ga
4-5$ M$_\odot$). We suggest that the water maser emission in IRAS
16552-3050 could be tracing a rapidly precessing bipolar
jet.

\end{abstract}

\keywords{masers -- stars: AGB and post-AGB -- stars: individual (IRAS
  16552$-$3050) -- stars: mass loss -- stars: winds, outflows}

\section{Introduction}

Water fountains are evolved Asymptotic Giant Branch (AGB) and young
post-AGB stars that show water maser emission with velocity
separations in their spectral features $\ge$100~km~s$^{-1}$
\citep{likkel88}. Some of them have been observed with radio
interferometric techniques and all show bipolar distributions of their
H$_2$O maser emission, indicating the presence of jets with extremely
short dynamical ages of 5-100 yr
\citep{likkel88,boboltz05,boboltz07,imai02,imai04,imai07iau}. OH maser
emission has been detected in most sources catalogued as water
fountains and, in general, this emission exhibits lower velocities and
is spatially less extended than that of H$_2$O \citep[see, for
example][]{deacon07}. Water fountains are evolutionarily located
between two phases of stellar evolution, AGB stars and planetary
nebulae (PNe), when mass-loss properties, among others, change
dramatically. While mass ejection during the AGB phase is spherical,
most PNe exhibit elliptical, bipolar, or multipolar morphologies.
These morphologies have been attributed to the action of collimated
outflows on the previously expelled spherical AGB shell
\citep{sahai98}. Therefore, water fountains very probably trace the
onset of non-spherical and highly collimated mass ejection.
Understanding the transformation of an AGB star into a PN relies on a
precise knowledge of the properties and evolution of water fountains.
Unfortunately, very few water fountains are known at present (11
sources reported as of March 2008, see \citealt{imai07iau}), so that
adding new members to this important class of objects represents a
valuable progress.

Recently, in a single-dish search for H$_2$O masers in evolved stars,
\citet{suarez07} found water maser emission toward IRAS\,16552$-$3050
(hereafter IRAS16552), with a maximum separation among its velocity
components of $\sim$170~km~s$^{-1}$. IRAS16552 was first proposed to
be a candidate post-AGB star by \citet{preite88}, based on its IRAS
colors. A possible optical counterpart of the IRAS far-infrared
source was classified by \citet{hu93} and \citet{suarez06} as a star
of spectral type K9III and K0I, respectively.

In this paper, we present Very Large Array (VLA) observations, carried
out to confirm the association of this high-velocity water maser
emission with the IRAS source, and to determine the spatial
distribution of the masers. Our results allow us to include IRAS16552
as a bona fide member of the water fountain class.  In Sec.~\ref{obs}
we describe the performed observations. Sec.  \ref{results} presents
the obtained results, and Sec. \ref{discussion} a discussion on the
properties of IRAS16552 and a comparison of this source with other water fountains.
Finally, in Sec. \ref{conclusions} we summarize the main conclusions
of this work.

\section{Observations and data reduction}
\label{obs}

We have observed the $6_{16}-5_{23}$ maser transition of the water
molecule (rest frequency $= 22235.08$ MHz) using the VLA of the
National Radio Astronomy Observatory (NRAO)\footnote{The National
  Radio Astronomy Observatory is a facility of the National Science
  Foundation operated under cooperative agreement by Associated
  Universities, Inc.} in two separate epochs. In the first one (2006
March 22 and 26), we used the A configuration of the VLA, and observed
the right circular polarization (1 IF mode) over a bandwidth of 3.11
MHz sampled with 255 channels (thus providing a spectral resolution of
1.22 kHz, or 0.16 km~s$^{-1}$). To be able to cover the whole range of
the water maser emission detected by \citet{suarez07} toward
IRAS16552, we successively centered the bandwidth at six different
velocities with respect to the Local Standard of Rest (LSR): $-70$,
$-30$, and $+10$ km~s$^{-1}$ on 2006 March 22, and $+50$, $+85$, and
$+120$ km~s$^{-1}$ on 2006 March 26. Therefore, we covered the water
maser emission of IRAS16552 on a velocity range from $-91$ to $+141$
km~s$^{-1}$ (although not simultaneously). The phase center for these
observations was set at R.A.(J2000) = $16^h58^m27.84^s$, Dec.(J2000) =
$-30^\circ 55'06.7''$, which is within $1''$ from the position of the
source given in \citet{suarez06}.

In the second epoch, carried out on 2007 December 24, we used the B
configuration of the VLA and the 4 IF mode, with a bandwidth of 3.08
MHz sampled with 63 channels for each IF (48.83 kHz, or 0.7 km
s$^{-1}$ spectral resolution). Two IFs were used to observe right and
left circular polarizations (RCP and LCP) centering the bandwidth at
$v_{\rm LSR} = -60$ km~s$^{-1}$, while the other two IFs (RCP and LCP)
were centered at $v_{\rm LSR} = +89.6$ km~s$^{-1}$. Based on the
results of the first epoch, we set the phase center of these
observations at R.A.(J2000) = $16^h58^m27.3^s$, Dec.(J2000) =
$-30^\circ 55'08''$, to include all observed maser components within $1''$
from this center.

For both epochs, source J1331+305 was used as the flux calibrator
(assumed flux density $\simeq 2.54$ Jy), J1229+090 was the bandpass
calibrator (bootstrapped flux density $\simeq 20.5\pm 1.1$, $26.4\pm
0.4$, and $25.2\pm 0.4$ Jy on 2006 March 22, 2006 March 26, and 2007
December 24, respectively) , while J1626-298 was the phase calibrator
(bootstrapped flux density $\simeq 1.00\pm 0.09$, $1.72\pm 0.03$, and
$2.30\pm 0.04$ Jy on 2006 March 22, 2006 March 26, and 2007 December
24, respectively). Maps were obtained with robust weighting of
visibilities (Robust parameter = 0) and deconvolved using the CLEAN
algorithm. The resulting synthesized beams were $\simeq 0.20''\times
0.07''$ and $\simeq 0.7''\times 0.25''$ for the first (A
configuration) and second (B configuration) epochs, respectively, with
their major axes oriented nearly north-south.

Data were self-calibrated using a strong maser feature, and applying
the obtained phase and amplitude correction to the rest of the
channels for each data set. In the first epoch, the maser components
detected in each day (2006 March 22 and 26, i.e., blue and redshifted
emission, respectively) could be aligned to a common reference
position, using spectral features that were present in the overlapping
region between each individual observation. However, the data taken in
different days could not be aligned since there was no common spectral
feature in the velocity ranges covered. Therefore, the 2$\sigma$
relative positional uncertainty between the maser features observed on
the same day in the first epoch is $\la 0.001''$ (estimated as $2\sigma
\simeq \frac{\Delta \theta}{\rm SNR}$, where $\Delta \theta$ is the
size of the primary beam and SNR is the signal to noise ratio), while
the positional uncertainty among the emission observed in different
days is $\simeq 0.1''$. For the second epoch, since all data were
taken simultaneously, the self-calibration corrections obtained for
the strongest maser feature was applied to all channels. The relative
positional accuracy among all features in that second epoch was $\la
0.005''$

\section{Results}
\label{results}

\begin{figure}
\includegraphics[scale=0.7]{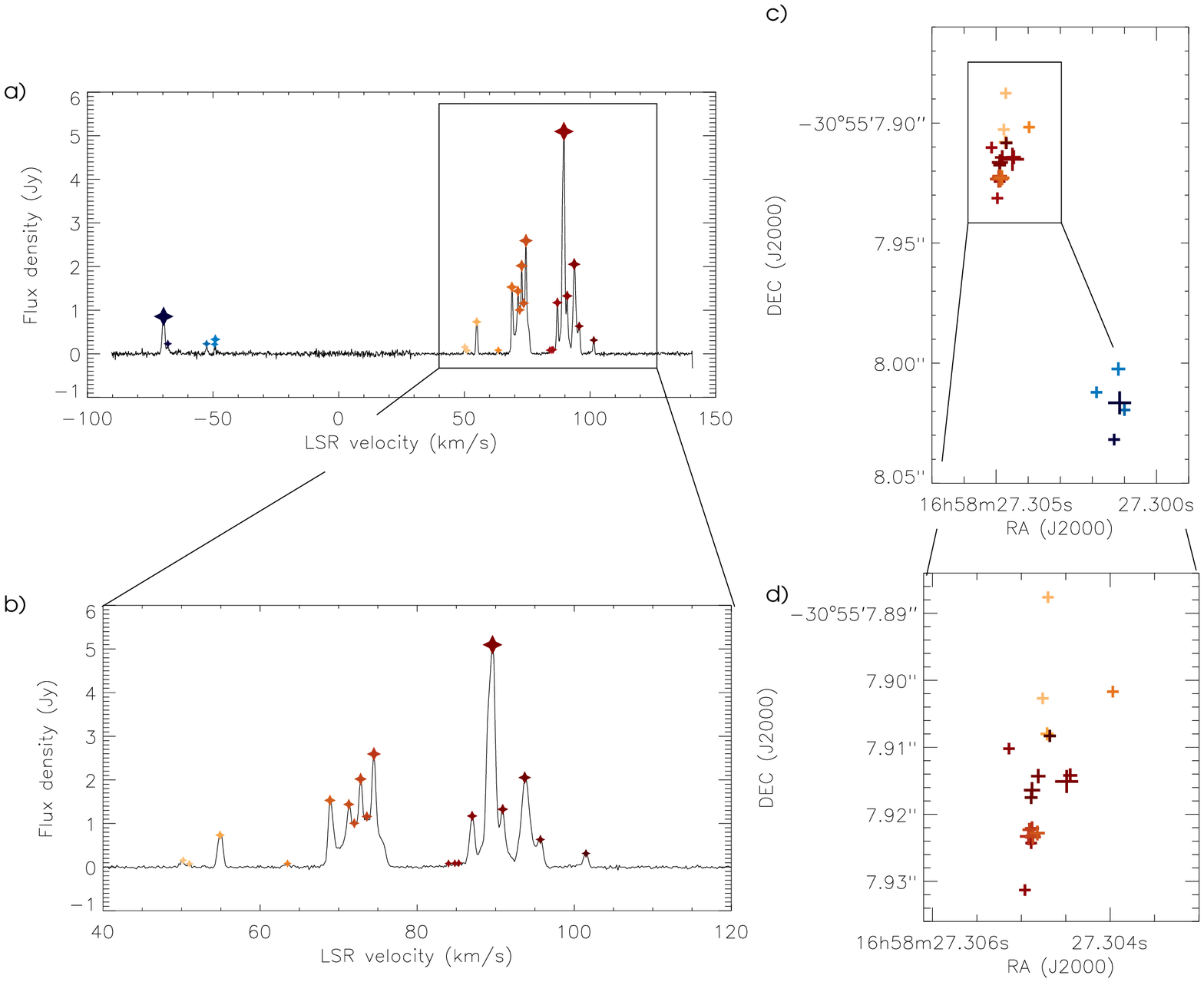}
\caption{(a) Spectrum of the water masers towards IRAS 16552$-$3050,
  observed on 2006 March. The different maser components are
  identified by colored stars. The color code is used to identify the
  same components in the rest of the panels. (b) Close up of the
  redshifted part of the spectrum. (c) Spatial distribution of the
  peaks of the maser components. Note that the coordinates in the
  figure should be considered only as nominal. The error in absolute
  position is $\simeq 0.1"$ (see Sec.~\ref{obs}). (d) Close up of the
  spatial distribution, showing only the redshifted components. The
  symbol size increases with flux density.}
\label{ini}
\end{figure}

The spectrum and the spatial distribution of the water maser
components obtained on the first epoch are shown in Fig.~\ref{ini}. By
``maser components'' we mean individual intensity peaks in the
spectrum. The positional information has been obtained by fitting
elliptical gaussians to the maser emission, only in the channels in
which a spectral peak is present. The total velocity span of the maser
emission is $\simeq$170~km~s$^{-1}$. Two groups of components are
evident in the spectrum: the redshifted group, from 50 to
110~km~s$^{-1}$, and the blueshifted one, from $-$40 to
$-$80~km~s$^{-1}$. The red components are more numerous and more
intense than the blue ones, with the most intense peak reaching
$\simeq 5.1$~Jy. The two groups appear spatially separated by $\simeq
0.11''$, forming a bipolar distribution, with the red components
clustered at the NE and the blue components clustered at the SW.
However, as explained in Sec.~\ref{obs}, while the spatial
distribution of components within either group is very reliable
(relative positional errors $\le 0.001''$), the relative position
between both groups has an error of $\simeq 0.1''$ (since they were
observed in different days), which is on the order of the apparent
separation. Therefore, the bipolar distribution of the maser emission
observed in the first epoch may not have been real.

\begin{figure}
 \includegraphics[scale=0.7]{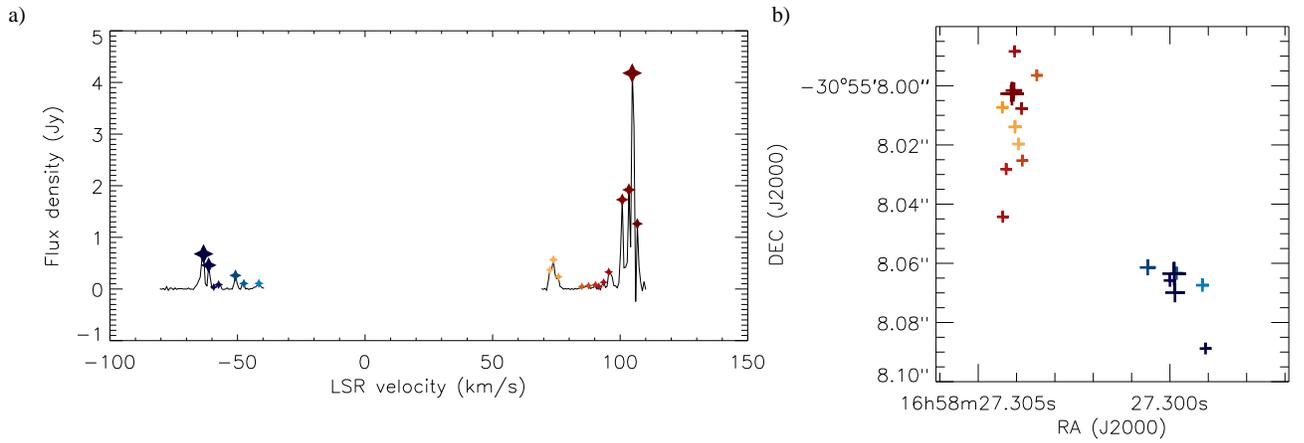}
 \caption{a) Spectrum of the water masers towards IRAS 16552$-$3050,
   observed on 2007 December. The different maser components are
   identified by colored stars. The color code is used to identify the
   same components in panel b. (b) Spatial distribution of the peaks
   of the maser components. Note that the coordinates in the figure
   should be considered only as nominal. The error in absolute
   position is $\simeq 0.1"$ (see Sec.~\ref{obs}). The symbol size
   increases with flux density.}
 \label{dec07}
 \end{figure}

Fig.~\ref{dec07} shows the spectrum and the spatial distribution of
the water maser emission obtained in the second epoch. These
observations did not cover the whole velocity range in which emission
is present. They were set up to simultaneously observe red- and
blueshifted components, in order to confirm the bipolarity. In these
data, the NE-SW bipolar distribution is present, with a separation of
$\simeq 0.08''$ between the centroids of the red and blue groups
along position angle (PA) $\simeq 44\deg$. The separation and PA are
both more reliable for this second epoch (relative position error
between features $\simeq 0.005''$), although the internal distribution
of the components in either group is more accurate in the first epoch.
Therefore, water maser emission from IRAS16552 seems to trace a
bipolar outflow with a projected size of $\simeq 0.08''$, and
velocity along the line of sight of $\simeq 85$ km~s$^{-1}$ (half of
the total velocity span).  Our results confirm IRAS16552 as a water
fountain with a bipolar distribution of its water maser emission. It
also confirms the trend of bipolarity in all water fountains studied
up to now at high angular resolution with radio interferometry,
strongly suggesting that water masers in all water fountains are
produced in collimated, bipolar jets.

The centroid position of the individual water maser components has
absolute coordinates R.A.(J2000) = $16^h58^m27.30^s$, Dec(J2000) =
$-30^{\circ}55'8.0''$ (absolute position error $\simeq 0.1''$,
compatible in both epochs).  Assuming a bipolar configuration for the
jet, we expect the powering source to be located near that position.
However, this centroid is $\simeq 16''$ away from the post-AGB star
identified at optical wavelengths by \citet{suarez06}, indicating that
the water maser emission is associated with a different evolved star
within the field.  Note that there is an error in the possition
given in Table 3 of \citet{suarez06}, since this position does not
correspond to the K0I star marked in the identification chart and
whose spectrum is shown in that paper, but to a nearby source (2MASS
J16582776-3055062). The true coordinates for the K0I star identified
in \citet{suarez06} are R.A.(J2000) = $16^h58^m28.53^s$, Dec(J2000) =
$-30{\deg}55'09.3''$.

We have searched the DSS plates and the 2MASS catalog for an optical
and/or near IR counterpart at the position of the outflow source,
but although several optical and near-IR sources lie within the IRAS
ellipse error, none of them is coincident with the position of the
maser emission. The closest 2MASS source, J16582725-3055108, does not
coincide within the errors (0.5$''$ for 2MASS) with the maser
position, since it is located at 2.3$''$ from the centroid position.
This implies that the outflow source must be heavily obscured by its
circumstellar envelope. From these results, it is plausible that the
IRAS point source actually traces this obscured source that is
powering the masers, but it is not related to the optical source. In
this case, the IRAS colors will classify IRAS16552 as a post-AGB star,
a classification also supported by its low IRAS variability. Its
likely post-AGB nature, together with its association with
high-velocity bipolar water-maser emission further indicates that
IRAS16552 belongs to the water fountain class.

As for the spatial distribution of water masers within each group, the
redshifted (NE) cluster of masers is oriented almost along the
north-south direction and extends $\simeq$ 0.04$''$-0.05$''$ in both
epochs. The SW group is also oriented almost north-south in the
first epoch and extends $\simeq$ 0.03$''$. The distribution in the
second epoch looks somewhat different, with an orientation along PA
$\sim$ 45$^{\circ}$, but given the relative errors (see Sec.
\ref{obs}), the distribution in this second epoch is less reliable. In
any case, it seems clear that, if we assume that the line joining the
red- and blueshifted groups traces the jet direction, the maser
emission within each group is not elongated along this jet.

As shown in Figs.~\ref{ini} and \ref{dec07}, the maser components
within the redshifted cluster tend to be grouped according to their
velocities, although the apparent grouping seems to be different in
the two epochs. In the first epoch, the least redshifted group, with
$v\leq 60$~km~s$^{-1}$ is located at the north, almost all the
components of the intermediate-velocity group ($60\leq v \leq
80$~km~s$^{-1}$) are located at the south, and the most redshifted
components ($v\geq 80$~km~s$^{-1}$) are situated in between the other
two groups. In the second epoch, the group centered $\simeq 75$
km~s$^{-1}$ is north of the one centered $\simeq 90$ km~s$^{-1}$,
which is the opposite distribution to the one in the first epoch. We
note again that in the second epoch we did not observe the same
velocity range as in the first one, and the positional error among the
components of each group is larger in the second one ($\simeq 0.005''$
as compared to $\simeq 0.001''$). In the SW group, no clear
tendency in the distribution of radial velocities is seen in either
epoch.

\section{Discussion}
\label{discussion}

\subsection{A comparison of IRAS16552 with other water fountains}
\label{comparison}

The absence of near IR counterpart in the Two Micron All Sky Survey
(2MASS) catalog \citep{skrutskie06} does not seem to be an uncommon
characteristic in the water fountains. Of the 11 water fountains
reported so far \citep[those listed in][]{imai07iau} all those covered
by the Midcourse Space Experiment (MSX) catalog \citep{egan03} (only
IRAS16552 was not included in the MSX survey) have a mid-infrared
counterpart. However, we have found 2MASS counterparts (coincident
with MSX sources within the errors) for only 5 of them. We did not
find any significant relationship between the presence of a near IR
counterpart and the characteristics of the water masers, such as
number of maser components, dynamical age, or velocity span. The still
small number of water fountains known is an obvious hindrance for this
type of comparison. However, the fact that about half of the water
fountains are not detected in the near IR gives us hints about the
characteristics of their envelopes compared to those of the rest of
AGB and post-AGB stars. Both AGB and post-AGB stars are usually
detected in the near IR, although a large fraction of AGB stars, and
some post-AGB stars are not detected in the optical \citep{suarez06}.
This suggest that the central stars of water fountains are surrounded
by thicker envelopes than in the rest of AGB and post-AGB stars. This
could be explained if the evolution into the post-AGB stage is faster
in water fountain sources, and therefore, these stars are relatively
massive ($\ga 4-5$ M$_\odot$).  Nevertheless, a study of the near IR
properties of water fountains, with a higher sensitivity than the data
in the 2MASS catalog may yield the detection of weak counterparts,
provided that an accurate position of the water maser emission is
determined with radio interferometric observations.

\begin{figure}
 \includegraphics[scale=1.]{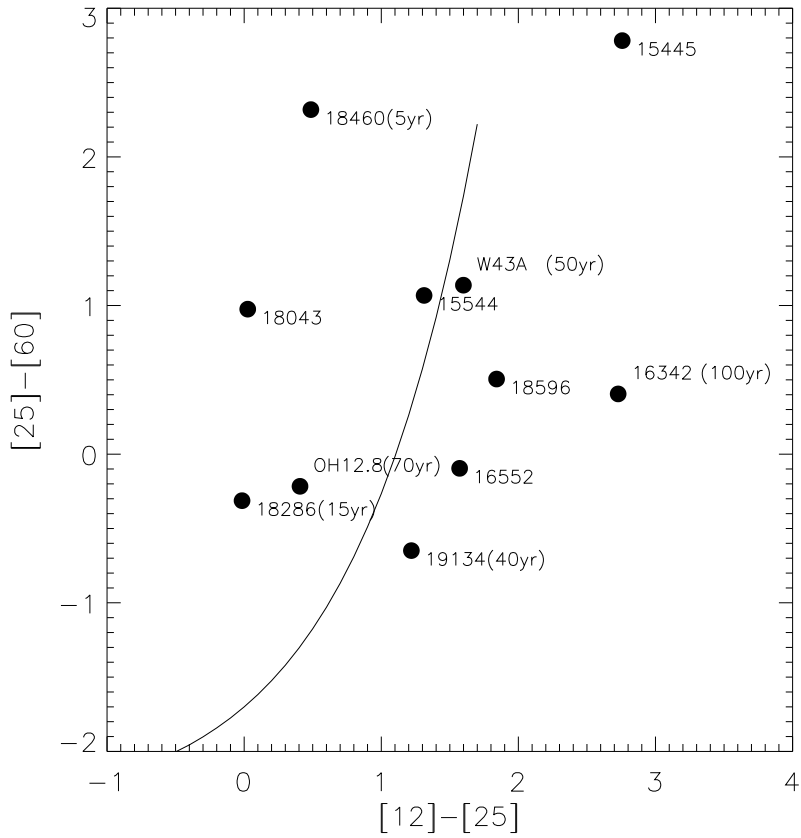}
 \caption{Position of the 11 water fountains known up to now in the
   IRAS two color diagram \citep[see][and references
   therein]{imai07iau}. The dynamical age is shown between parentheses
   for those objects for which it has been calculated. The curve in
   the plot is the one defined by \citet{bedijn} where the AGB stars
   tend to be located.}
 \label{coloriras}
 \end{figure}

In order to place IRAS16552 in the evolutionary framework of water
fountains, we have plotted all such sources known on an IRAS
two-color diagram (Fig.~\ref{coloriras}), where we also show their
dynamical ages (for those known). It is interesting that water
fountains tend to be distributed in this IRAS color-color diagram on
the same region as post-AGB stars \citep{suarez06}, while in this
region, few AGB stars are located \citep{jimenezesteban06}.  This is
certainly reasonable, given that water fountains seem to be post-AGB
or evolved AGB stars. However, it is potentially important, since it
suggests that IRAS colors could be useful as diagnostics to select
potential candidate sources in future searches for water fountains
among the most extreme AGB stars.

\begin{figure}
 \includegraphics[scale=1.]{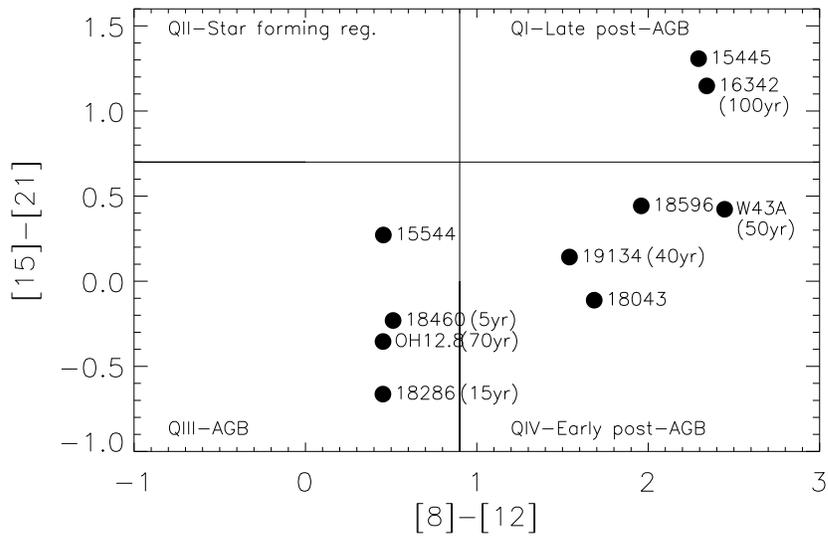}
 \caption{MSX two-color diagram of the 10 water
   fountains known up to now that have been observed with this
   satellite. The four quadrants are those defined in
   \citet{sevenster02a} and described in the text.}
 \label{colormsx}
 \end{figure}

To try to establish the evolutionary stage for each of the water
fountains we have used the [8]$-$[12] {\it vs} [15]$-$[21] MSX
two-color diagram (see Fig~\ref{colormsx}). \citet{sevenster02a} has
studied the position of OH-maser-emitting AGB and post-AGB sources in
this diagram. The diagram has been divided in four quadrants (QI, QII,
QIII, QIV), with classical AGB stars located in QIII (low left), early
post-AGB stars in the QIV (low right) and more evolved objects in the
QI, while QII contains star forming regions. The two water fountains
with the shortest dynamical age are located in region III, and the one
with the longest in region I, which agrees with Sevenster's
evolutionary distribution. Estimates of the dynamical ages of the rest
of water fountains will be helpful to confirm this trend. Moreover,
the objects with visible lobes (IRAS 16342$-$3814 and IRAS 19134+2131) are in
the QIV and QII region, always with [8]$-$[12]$>$0.9. According to
\citet{sevenster02a}, the transition from [8]$-$[12]$<$0.9 to
[8]$-$[12]$>$0.9 implies the transition from AGB to post-AGB stars.

The water maser components in water fountains seem to show always a
spatial bipolar layout, although their distribution is rather
different from one object to another. Also the number of components is
very variable from one source to another. There are sources as
OH12.8$-$0.9 which show only 2 or 3 water maser components with very
similar velocities at the tips of each lobe
\citep{boboltz05,boboltz07}, and sources as IRAS 18286$-$0959 that
show multiple components with significant velocity dispersion within
each lobe, and also components located close to the center of the
maser distribution, with intermediate velocities between those of the
most extreme components \citep{imai07iau}. In the case of IRAS16552,
it shows at least 15 components with a spread in velocities within
each lobe of $\simeq 25-50$~km~s$^{-1}$, and an absence of equatorial
masers. These spectral and spatial characteristics make it similar to
other water fountains as IRAS 16342$-$3814 \citep{morris03}, or IRAS
19134+2131 \citep{imai04,imai07}.

We note that single-dish observations of the OH maser line at 1612 MHz
did not show any maser emission in IRAS16552
\citep{telintel91arti,hu94}, with 1$\sigma$ rms of 0.2, and 0.1 Jy,
respectively. The presence or not of OH maser emission, as well as its
kinematical and spatial characteristics is of potential interest to
further characterize water fountains. The presence of OH maser
emission is a common characteristic in almost all water fountains.
Apart from IRAS16552, the only other known water fountain in which OH
seems not to be present is IRAS 19134+2131 \citep{lewis87,hu94}. The
OH masers detected are in general less extended and have a smaller
range of velocities than the H$_2$O masers, as is the case of IRAS
18043$-$2126 \citep{sevenster01, deacon07}, OH12.8$-$0.9
\citep{boboltz05}, W43A \citep{imai02}. This suggest that the OH
masers tend to trace the former AGB shell, rather than the later
bipolar mass-loss traced by H$_2$O. However, the cases of IRAS
15445-5449 and IRAS 15544-5332 \citep{deacon04, deacon07} do not
follow this trend, since OH maser emission in these sources have been
found outside the velocity range of the H$_2$O masers. IRAS 16342$-$3814
can be considered as another exception since although the OH masers
have a smaller velocity range than the H$_2$O masers (70~km~s$^{-1}$
vs 160~km~s$^{-1}$; \citealt{sahai99,zijlstra01,likkel88,morris03}), they are located in
the bipolar jets and have a velocity that does not correspond to the
AGB shell.

\subsection{The orientation of the water maser distribution}

As we mentioned in Sec. \ref{results}, the water maser emission in
IRAS16552 is not aligned with the apparent jet direction. Such
misalignment is also common in other sources, as IRAS 19134+2131
\citep{imai04,imai07} and W43A \citep{imai02}. Those cases can be
explained as the effect of ballistic corkscrew jets
\citep{imai07iau,vlemmings06}, where the masers are emitted in a
ballistic way along the direction of a jet that precesses with time.
This precession could be due to the presence of a binary stellar
companion or a massive planet. These companions would also account for
the magnetic fields found in sources as W43A \citep{vlemmings06}.
Nevertheless, no direct evidence for a companion has been found so far
in any water fountain.

IRAS16552 is an extreme case in which the water maser components are
oriented almost perpendicular to the radial direction from the central
source (assumed to be at the center of the distribution of the
components, see Sec. \ref{results}). An almost perpendicular
distribution is also seen in IRAS 16342$-$3814 \citep{morris03} and OH
12.8-0.9 \citep{boboltz05}, which has been attributed to the masers
tracing a shock front. For IRAS16552, such an interpretation would
imply a rather large opening angle of the jet ($\simeq 70^\circ$). A
possible alternative explanation is that the water masers in IRAS16552
trace a rapidly precessing/rotating bipolar jet. If the
precession/rotation period is very short, as compared with the
expansion timescale determined by the jet velocity, the bipolar jet
(or the shocked regions in the neutral envelope) will describe
point-symmetric structures oriented almost perpendicular to the
outflow direction. At least qualitatively, this scenario may account
for the observations of IRAS16552, specially for the first epoch, in
which the distribution of masers within each group has been determined
with a higher accuracy. 

Remarkably, many PNe show collimated structures oriented almost
perpendicular to the radial direction from the central star. This is
the case of bipolar and elliptical PNe that exhibit a point-symmetric
intensity distribution in the shell as, for instance, Hb5
\citep{corradi93} and Cn3-1 \citep{miranda97}. More interesting is the
case of components DD' in NGC6543 \citep{balick04}, which are dense
(10$^{4}$ - 10$^{5}$ cm$^{-3}$) filaments that have been interpreted
as due to a rapidly precessing bipolar jet \citep{miranda92}.
Given that water fountains are probably related to relatively massive
stars (Sec. \ref{comparison}), their evolution through the post-AGB
phase should be very fast. The fingerprints of rapidly precessing,
collimated post-AGB outflows, such as the one proposed for IRAS16552,
may still be present in the envelope when the central star ionizes the
nebula and could show up as point-symmetric ionized filaments in PNe,
similar to the components DD' in NGC6543.

\section{Conclusions}
\label{conclusions}

In this paper we have presented VLA observations of H$_2$O masers at
22 GHz towards IRAS 16552$-$3050.  Our main conclusions are as follow:

\begin{itemize}

\item We have found that the water maser emission in IRAS 16552$-$3050
  span a velocity range of $\sim$170~km~s$^{-1}$ and shows a bipolar distribution of
  $\simeq$ 0.08'' in size. The properties of the water maser
  emission and the likely post-AGB nature of the central source allow
  us to confirm IRAS 16552$-$3050 as a member of the class of ``water
  fountain'' evolved objects.

\item Our result confirms the trend of bipolarity in all water
  fountains known up to now, strongly suggesting that water masers in
  these sources are produced in bipolar, collimated jets.

\item The water maser emission in IRAS 16552$-$3050 does not seem to
  be associated with the optical source identified by \citet{hu93} and
  \citet{suarez06}, but with a different evolved object, for which
  there is no optical nor near-IR counterpart known.

\item About half of the water fountains known do not have a near-IR
  counterpart in the 2MASS catalog. This suggest that water fountains
  are surrounded by thicker envelopes than the rest of AGB and
  post-AGB stars (which usually have near-IR counterparts), probably
  implying that they are relatively massive objects ($\ga 4-5$
  M$_\odot$).

\item The distribution of the water maser emission in the red- and
  blueshifted group is almost perpendicular to the proposed jet
  direction. This might be explained if the maser emission traces a
  rapidly precessing/rotating bipolar jet. A later evolution could
  give rise to point-symmetric filaments observed in some PNe, which
  appear oriented almost perpendicular to the radial direction from
  the central star.

\end{itemize}

\acknowledgments

The authors with to thank Drs. G. Anglada and M. Osorio for useful
discussions. JFG and LFM acknowledge support from Ministerio de
Ciencia e Innovaci\'on (Spain), grants AYA 2005-08523-C03 and AYA
2005-01495, respectively, co-funded with FEDER funds. OS, JFG, and LFM
are also supported by Consejer\'{\i}a de Innovaci\'on, Ciencia y
Empresa of Junta de Andaluc\'{\i}a. This publication makes use of data
products from the Two Micron All Sky Survey, which is a joint project
of the University of Massachusetts and the Infrared Processing and
Analysis Center/California Institute of Technology, funded by the
National Aeronautics and Space Administration and the National Science
Foundation.


\clearpage

\end{document}